\def\rfr#1{Eq. (\ref{#1})}

\def\leti{Lense--Thirring}
\def\Rfr#1{Eq. (\ref{#1})}

\def\bar{\begin{eqnarray}}
\def\ear{\end{eqnarray}}

\def\eqi{\begin{equation}}
\def\eqf{\end{equation}}
\def\eqia{\begin{eqnarray}}
\def\eqfa{\end{eqnarray}}
\def\rp#1#2{{#1\over#2}}

\def\lb#1{\label{#1}}

\def\oc2{$\mathcal{O}(c^{-2})$}

\documentclass[11pt]{article}
\usepackage{amsmath,amsthm,amscd,amssymb}
\usepackage{latexsym}
\usepackage{graphicx,epsfig}

\begin{document}

\noindent{\bf \LARGE{The challenge to reliably measure the general
relativistic Lense-Thirring effect with a few percent accuracy}}
\\
\\
\\
{Lorenzo Iorio}\\
{\it Dipartimento Interateneo di Fisica dell' Universit${\rm
\grave{a}}$ di Bari
\\Via Amendola 173, 70126\\Bari, Italy
\\e-mail: Lorenzo.Iorio@libero.it}

\begin{abstract}
In this paper we critically analyze the so far performed and
proposed tests for measuring the general relativistic
Lense-Thirring effect in the gravitational field of the Earth with
some of the existing accurately tracked artificial satellites. The
impact of the 2nd generation GRACE-only EIGEN-GRACE02S Earth
gravity model and of the 1st CHAMP+GRACE+terrestrial gravity
combined EIGEN-CG01C Earth gravity model is discussed. The role of
the proposed LARES is discussed as well.
\end{abstract}

\section{Introduction}

The Einstein's General Theory of Relativity (GTR), in its
weak-field and slow-motion approximation (Soffel 1989) valid
throughout the Solar System, predicts, among other things, that
the rotation of a body of mass $M$ induces a so called
`gravitomagnetic' component of its gravitational field which acts
on a test particle orbiting it with a non central,
velocity-dependent force analogous to the Lorentz force of the
Maxwellian electromagnetism. As a consequence, the longitude of
the ascending node $\Omega$ and the argument of pericentre
$\omega$ of its orbit undergo tiny secular precessions \eqi
\dot\Omega_{\rm LT}=\rp{2GJ}{c^2a^3(1-e^2)^{3/2}}, \dot\omega_{\rm
LT}=-\rp{6GJ\cos i}{c^2a^3(1-e^2)^{3/2}},\eqf where $G$ is the
Newtonian constant of gravitation, $J$ is the proper angular
momentum of the central body, $c$ is the speed of light, $a,e$ and
$i$ are the semimajor axis, the eccentricity and the inclination,
respectively, of the test particle's orbit. It is the
Lense-Thirring effect (Lense and Thirring 1918).

Recent years have seen increasing efforts devoted to the
measurement of this kind of post-Newtonian gravitomagnetic effect
in the gravitational field of the Earth by means of the analysis
of the Satellite Laser Ranging (SLR) data to the existing LAGEOS
($a=12270$ km, $i=110$ deg, $e=0.0045$) and LAGEOS II ($a=12163$
km, $i=52.65$ deg, $e=0.014$) geodetic satellites (Ciufolini 2004;
Ciufolini and Pavlis 2004; Iorio and Morea 2004; Iorio 2004) for
which the Lense-Thirring effect amounts to a few tens of
milliarcseconds per year (mas yr$^{-1}$). Alternative approaches
including also the other existing SLR satellites, with particular
emphasis on Ajisai, and the altimeter Jason-1 satellite have also
been proposed (Iorio 2002; Iorio and Doornbos 2004; Vespe and
Rutigliano 2004). The originally proposed LARES mission (Ciufolini
1986; 1998; Iorio et al. 2002; Iorio 2003a), which involves the
launch of another SLR target whose data should be analyzed
together with those from LAGEOS and LAGEOS II, has recently been
investigated in the context of the relativity-dedicated OPTIS
mission (Iorio et al. 2004; L\"{a}mmerzahl et al. 2004).

The major sources of systematic errors in such kind of
measurements are due to the aliasing classical secular precessions
(Iorio 2003b) induced by the even zonal harmonic coefficients
$J_{\ell}, \ell=2,4,6,...$ of the multipolar expansion of the
terrestrial gravitational potential, called geopotential (Kaula
1966), and to the non-gravitational perturbations induced, e.g.,
by the direct solar radiation pressure, the Earth albedo, the
Earth infrared radiation, the solar Yarkovsky-Schach effect, the
terrestrial Yarkovsky-Rubincam effect, the asymmetric reflectivity
(Lucchesi 2001; 2002; 2003; 2004; Lucchesi et al. 2004). The
non-gravitational forces especially affect the perigees of the
geodetic satellites of LAGEOS type, while the nodes are relatively
insensitive to them. The observables used or proposed are suitable
linear combinations of the orbital residuals of the rates of the
nodes and the perigees of various existing and proposed
satellites. Their main goal is to reduce the impact of the
systematic error due to the mismodelling in the static and the
time-varying parts of the geopotential's coefficients by
cancelling out as many even zonal harmonics as possible. On the
other hand, a reasonable compromise with the fact that the impact
of the non-gravitational forces has to be reduced as well must
also be obtained. It is important to note that the perturbations
of gravitational origin  have the same linear temporal signature
of the Lense-Thirring effect itself, or are even parabolic if the
secular variations $\dot J_{\ell}$ of the even zonal harmonics are
accounted for, while many of the non-gravitational forces (and all
the tidal perturbations (Iorio 2001)) are periodic. This means
that, for a given observational time span $T_{\rm obs}$, the
harmonic noises can be fitted and removed from the time series,
provided that their periods $P$ are shorter than $T_{\rm obs}$:
this is not possible for the linear and parabolic biases without
distorting the genuine relativistic linear signal of interest. So,
their impact on the performed measurement can-and must-only be
assessed as more accurately and reliably as possible. Another
important point to be noted is that the observables should also be
chosen in order to reduce the effect of the a priori `memory
imprint' of GTR on the background reference Earth gravity models
adopted in the analysis. Indeed it could drive the outcome of the
tests just towards the expected result compromising their full
reliability.

In this paper we wish to critically discuss the performed and
proposed attempts to detect the Lense-Thirring effect in view of
the progress in our knowledge of the classical part of the
terrestrial gravitational field due to the dedicated CHAMP (Pavlis
2000) and, especially, GRACE (Ries et al. 2003a) missions.
\section{The node-node-perigee tests}
The first attempts to measure the Lense-Thirring effect were made
with the following combination of the orbital residuals of the
rates of the nodes of LAGEOS and LAGEOS II and the perigee of
LAGEOS II (Ciufolini 1996) \eqi \delta\dot\Omega^{\rm LAGEOS
}+c_1\delta\dot\Omega^{\rm LAGEOS\ II}+c_2\delta\dot\omega^{\rm
LAGEOS\ II}\sim \mu60.2,\lb{ciufform} \eqf where $c_1=0.304$,
$c_2=-0.350$ and $\mu$ is the solved--for least squares parameter
which is 0 in Newtonian mechanics and 1 in
 GTR. The predicted relativistic signal is a linear trend with a
 slope of 60.2 mas yr$^{-1}$. The combination of \rfr{ciufform}
 cancels out $J_2$ and $J_4$ along with their temporal variations.

In (Ciufolini 2004) the results of the tests performed with
\rfr{ciufform} and the pre-CHAMP/GRACE EGM96 Earth gravity model
(Lemoine et al. 1998) are reported with a claimed total error of
20-25$\%$. In reality, this estimate is widely optimistic, as
pointed out by a number of authors (Ries et al. 2003b; Iorio
2003b; Iorio and Morea 2004; Vespe and Rutigliano 2004). Indeed,
in the EGM96 solution the retrieved even zonal harmonics are
reciprocally strongly correlated, so that a realistic evaluation
of the systematic error induced by them should be performed by
linearly adding the absolute values of the individual errors. In
this case a 1-$\sigma$ 83$\%$ error is obtained. Instead,
Ciufolini has used the full covariance matrix of EGM96 obtaining a
1-$\sigma$ 13$\%$ error which comes from a luckily correlation
between the uncancelled $J_6$ and $J_8$. The point is that such
covariance matrix has been obtained from a multidecadal analysis
of the data from a large number of SLR satellites among which
LAGEOS and LAGEOS II played an important role. The Lense-Thirring
tests, instead, have been conducted over time spans few years
long, so that nothing assures that the EGM96 covariance matrix
realistically reflects the correlations among the even zonal
harmonics during any particular relatively short time spans.

Also the impact of the non-gravitational perturbations on the
perigee of LAGEOS II has been underestimated. According to
Lucchesi (2002), their systematic error would amount to almost
28$\%$ over 7 years. Moreover, the effect of the Earth's penumbra
on it (Vespe 1999) has not been considered at all: it would amount
to 10$\%$ over 4 years. Finally, although mainly concentrated in
$J_2$ and $J_4$, the Lense-Thirring `imprint' is also presented in
EGM96 which is largely based just on the LAGEOS satellites which
have been used for measuring the gravitomagnetic effect.
\section{The node-node tests}
The notable improvements in our knowledge of the Earth's
gravitational field thanks to the CHAMP and GRACE missions have
allowed to look for alternative combinations capable of reducing
the total error. In (Iorio and Morea 2004; Iorio 2005a) the
following combination has been explicitly proposed\footnote{The
possibility of using only the nodes of LAGEOS and LAGEOS II in
view of the future benefits from the GRACE mission was put forth
in (Ries et al. 2003b) for the first time, although without
quantitative details.} \eqi\delta\dot\Omega^{\rm LAGEOS }_{\rm obs
}+k_1\delta\dot\Omega^{\rm LAGEOS\ II}_{\rm obs}\sim \mu_{\rm
LT}48.2,\lb{iorform}\eqf where $k_1= 0.546$ and 48.2 is the slope,
in mas yr$^{-1}$, of the expected gravitomagnetic linear trend.
\Rfr{iorform} dramatically reduces the systematic error due to the
non-gravitational perturbations to $\sim 1\%$ because it does not
include the perigee of LAGEOS II.

The assessment of the systematic error of gravitational origin is
rather subtle (Iorio 2005b) because of the fact that the
combination of \rfr{iorform} only cancels out $J_2$ along with its
temporal variations. Instead, $J_4, J_6, J_8,..$ do affect it. In
particular, the impact of $\dot J_4$ and $\dot J_6$, for which
large uncertainties still exist (Cox et al. 2003), may be a
limiting factor, especially over time spans many years long.
Indeed, as already pointed out, they would induce a parabolic
noise signal which could not safely be fitted and removed from the
time series without distorting the trend of interest as well.

Moreover, also the problem of the a priori GTR `memory' effect
would still be present even with the GRACE-based model. In fact,
GTR has not been modelled in the currently released
GeoForschungZentrum (GFZ, Potsdam) GRACE-based models (F.
Flechtner, GFZ team, private communication, 2004) like
EIGEN-GRACE02S (Reigber et al. 2005) and EIGEN-CG01C (Reigber et
al. 2004). GRACE recovers the low degree even zonal harmonics from
the tracking of both satellites by GPS and the medium-high degree
geopotential coefficients from the observed intersatellite
distance variations. From (Cheng 2002) it can be noted that the
variation equations for the Satellite-to-Satellite Tracking (SST)
range $\Delta\rho$ and range rate $\Delta\dot\rho$ of GRACE can be
written in terms of the in-plane radial and, especially,
along-track components $R,T, V_R, V_T$ of the position and
velocity vectors, respectively. In turns, they can be expressed as
functions of the perturbations in all the six Keplerian orbital
elements (see (10)-(11), (A4)-(A6), (A14)-(A16) and (A28)-(A30) of
(Cheng 2002) ). Now, the gravitomagnetic off-diagonal components
of the spacetime metric also induce short-periodic 1 cycle per
revolution (1 cpr) effects (Lense and Thirring 1918; Soffel 1989)
on all the Keplerian orbital elements, apart from the secular
trends on the node and the pericentre. This means that there is
also  a Lense-Thirring signature in all the other typical
satellite and intersatellite observables like ranges and
range-rates. It is likely that it mainly affects just the
low-degree even zonal harmonics to which the $J_2$-free
combination of \rfr{iorform} is sensitive.

In regard to the static part of the geopotential, the systematic
error induced by it amounts to $4\%$ (1-$\sigma$ upper bound)
according to the GRACE-only EIGEN-GRACE02S model and to $6\%$
(1-$\sigma$ upper bound) according to EIGEN-CG01C which combines
data from CHAMP, GRACE and terrestrial gravimetry. According to
the evaluations of Cox et al. (2003), the impact of $\dot J_4$ and
$\dot J_6$, which grows linearly in time by assuming that no
inversions in their rates of change occur, is of the order of
1$\%$ yr$^{-1}$ (1-$\sigma$).

Another point to be noted is that in the aforementioned GFZ models
$\dot J_2$ and $\dot J_4$ have not been solved for: instead, they
have been held fixed to given values obtained from long time
series of SLR data to the geodetic satellites among which LAGEOS
and LAGEOS II, again, played a relevant role. $\dot J_6$ is not
present at all. This means that in the recovered even zonal
harmonics an `imprint' from such secular variations is probably
present, so that in evaluating the total error of gravitational
origin it would be more conservative and realistic to linearly add
the effects due to $J_{\ell}$ and those due to $\dot J_{\ell}$.

The combination of \rfr{iorform} has been used for tests with real
data by Ciufolini and Pavlis (2004) over an observational time
span of 11 years with EIGEN-GRACE02S . They incorrectly attribute
the $J_2-$ free node-only combination to themselves by means of
(Ciufolini 1986) which, instead, has nothing to do with it.
Moreover, they claim a total error ranging from 5$\%$ (1-$\sigma$)
to 10$\%$ (3-$\sigma$), but, again, it seems to be too optimistic
and incorrectly evaluated (Iorio 2005b). E.g., the issues of the
time-varying part of the geopotential and of the Lense-Thirring
`imprint' have been completely neglected. Moreover,
root-sum-square calculations have been often ad-hoc used, when
caution would have advised to linearly sum, e.g., the various
errors of gravitational origin which can hardly be considered as
independent. More realistic evaluations including also the effects
of $\dot J_4$ and $\dot J_6$ points toward a
$15\%(1-\sigma)-45\%(3-\sigma)$ range error. Even by assuming the
unsupported $2\%$ claimed by Ciufolini and Pavlis for the
time-dependent gravitational part of the systematic error, a more
realistic evaluation of the total uncertainty yields a
$6\%(1-\sigma)-19\%(3-\sigma)$ interval. Finally, a scatter plot
obtained by using different Earth gravity models and different
observational time spans should have been produced.
\section{An alternative combination}
A way to reduce to systematic error due to the geopotential is, in
principle, to suitably combine the nodes of $N$ satellites so to
cancel out the first $N-1$ even zonal harmonics (Iorio 2002; Iorio
and Doornbos 2004; Vespe and Rutigliano 2004). This possibility is
very appealing because for, say, $N=4$ $J_2, J_4$ and $J_6$, along
with their temporal variations and a large part of the a priori
GTR `memory' effects would be canceled out. The main practical
problem is that the other existing SLR satellites are too low
(Starlette, Stella, Larets) or too high (ETALON1, ETALON2) with
respect to LAGEOS and LAGEOS II.

The use of the ETALON satellites ($a=25498$ km) would imply huge
coefficients of their nodes which would enhance the effect of the
uncancelled $\ell=2,m=1$ tesseral $K_1$ tide on the obtainable
combinations. Indeed, the periods of such orbital perturbations
are equal to the periods of the nodes which, for the ETALON
satellites, amount to tens of years. Moreover, the nominal
amplitudes of such semisecular signals are of the order of
thousands of milliarcseconds.

On the other hand, the lower satellites with $a\sim 7000$ km would
not pose problems in regard to the harmonic perturbations because
their periods amount to months or a few years; the major drawback
is represented by the uncancelled even zonal harmonics of higher
degree to which such satellites are much more  sensitive than
LAGEOS and LAGEOS II.

The SLR Ajisai satellite ($a=7870$ km) and the altimeter Jason-1
satellite ($a=7713$ km) lie in an intermediate position. Indeed,
it turns out that their nodes could be usefully combined, in
principle, with those of LAGEOS and LAGEOS II in order to reduce
the impact of the geopotential to the 1$\%$ level without the
uncertainties related to the $\dot J_{\ell}$ and the a priori
imprints of the background Earth gravity models to be used. In
(Iorio and Doornbos 2004) the following $J_2-J_4-J_6$-free
combination has been proposed \eqi\delta\dot\Omega^{\rm
L}+h_1\delta\dot\Omega^{\rm L\ II}+h_2\delta\dot\Omega^{\rm
Aji}+h_3\delta\dot\Omega^{\rm Jason}\sim\mu49.5,\lb{jason}\eqf
with \eqi h_1=0.347,\ h_2=-0.005,\ h_3=0.068.\lb{jasoncoef}\eqf
According to EIGEN-GRACE02S, the 1-$\sigma$ upper bound for the
systematic error due to the geopotential amounts to 2$\%$, while
it is 1.6$\%$ according to EIGEN-CG01C. Note that, in regard to
the Lense-Thirring effect, \rfr{jason} is sensitive to the even
zonal harmonics up to $\ell=20$: this allows for accurate and
reliable evaluations of their systematic error also in an
analytical way (Iorio 2003b). It is likely that the forthcoming
models will push the systematic error of gravitational origin
below the 1$\%$ level. In regard to the most insidious uncancelled
tidal perturbations like $K_1$ acting on the nodes of Ajisai and
Jason-1, their periods amount to almost half a year. The major
drawback of the combination of \rfr{jason} is represented by the
impact of the non-gravitational perturbations on Jason-1 which
should be modelled in a truly accurate dynamical way: indeed its
area-to-mass ratio, to which the non-conservative forces are
proportional, amounts to $\sim 2.7\times 10^{-2}$ m$^2$ kg$^{-1}$,
contrary to $7\times 10^{-4}$ m$^2$ kg$^{-1}$ of LAGEOS. However,
according to the evaluations in (Iorio and Doornbos 2004) they
should mainly have harmonic signatures with periods of the order
of 1 year. This is a very important feature because they could,
then, be fitted and removed from the time series over not too long
observational time intervals. Moreover, the small magnitude of the
coefficient $h_3$ which weighs the Jason's node would be helpful
in keeping the non-conservative forces within the few percent
level and in reducing the measurement errors. It should be pointed
out that, up to now, there are no long time series of the
out-of-plane cross track Keplerian orbital elements of Jason
available, contrary to the in-plane radial and along-track
components of its orbit due to its oceanographic and altimetric
use. Moreover, the current (radial) 1-cm accuracy in
reconstructing its orbit is obtained in a reduced-dynamic fashion
which would be unsuitable for Lense-Thirring tests. Also the
orbital maneuvers may affect the possibility of getting smooth
time series some years long, although they are mainly performed in
its orbital plane.
\section{The LARES mission}
In its originally proposed version the LAGEOS III/LARES satellite
is a SLR twin of LAGEOS which has to be launched in the same orbit
of LAGEOS, apart from the inclination whose nominal value is
$i=70$ deg (Ciufolini 1986) and the eccentricity whose nominal
value is $e=0.04$ (Ciufolini 1998). The observable is the simple
sum of the nodes of LAGEOS and LARES which would cancel, to a
certain degree of accuracy depending on the precision of the
launch and, consequently, on the quality and the cost of the
launcher, all the classical precessions of the geopotential, which
are proportional to $\cos i$ (Iorio 2003b), and would enforce the
Lense-Thirring total signature, which, instead, is independent of
$i$. However, according to the EGM96 Earth gravity model,
departures from the nominal inclination up to 1 deg would induce a
gravitational error of 10$\%$ (1-$\sigma$ upper bound). The more
recent EIGEN-CG01C model reduces this limit to 2$\%$. Moreover,
also the $\dot J_2, \dot J_4, \dot J_6 $, to which the sum of the
nodes is sensitive, would further corrupt the obtainable accuracy
over a time span of some years.

Later, in (Iorio et al. 2002), it was proposed to suitably combine
the node and the perigee of LARES with the nodes of LAGEOS and
LAGEOS II and the perigee of LAGEOS II in order to greatly reduce
the dependence on the unavoidable orbital injection errors, so to
somewhat relax the original very stringent requirements on the
required LARES orbital configuration. The systematic error due to
the remaining uncancelled even zonal harmonics, calculated with
the EGM96 Earth gravity model, was well below the 1$\%$ without
the uncertainties related to the $\dot J_{\ell}$.

It is important to note that, since LARES is totally passive as
LAGEOS and LAGEOS II and since its data must be combined together
with those of its already orbiting twins, the level of the
obtainable accuracy in measuring the \leti\ effect with it would
be set by the non-gravitational perturbations, i.e. $\sim 1\%$.
Then, it would be  unnecessary  to push the systematic error of
gravitational origin much below the 1$\%$ level. This
consideration, together with the precision reached by the
present-day (and future) Earth gravity models, allow for a much
greater freedom in choosing the orbital configuration of LARES
with respect to its original configuration. In particular, it
would be possible to greatly reduce the costs of the launch by
inserting LARES in a much lower orbit with respect to that of
LAGEOS. E.g., an orbital configuration like that of Jason-1 would
be well suited: the combination of \rfr{jason}, with Jason's node
replaced by the node of a low-altitude LARES with its orbital
parameters, would easily reach a $\sim 1\%$ level of accuracy. It
is also possible to show that with a three-node combination of the
nodes of the LAGEOS satellites and of a relatively low-altitude
LARES ($a\sim 8000$ km) a satisfactory gravitational error of less
than 1$\%$ ($1-\sigma$ upper bound obtained with EIGEN-CG01C)
could be achieved.

Of course, if the implementation of the combination of \rfr{jason}
with Jason-1 will be really feasible and/or the uncertainties
related to the use of \rfr{iorform} will be reduced in some ways,
the cost of an entirely new dedicated mission which would allow to
only reach a $\sim 1\%$ measurement of the Lense-Thirring effect
by means of a passive SLR satellite could be judged
unjustified\footnote{The situation with the OPTIS mission is
different because the measurement of the Lense-Thirring effect
would be one of its many other relativistic tasks which could  be
implemented with comparatively small modifications of the original
concept.}. The launch of at least two drag-free satellites
could, in principle, be better justified because it would allow a
really notable improvement in the error budget thanks to the
active reduction of the non-gravitational perturbations. Note also
that an active compensation of the non-conservative forces would
also make feasible the use of their perigees along with their
nodes without resorting to the passive LAGEOS and LAGEOS II; with
the supplementary plane option it would also be possible to
measure the difference of the perigees (Iorio and Lucchesi 2003).
The forthcoming CHAMP/GRACE Earth gravity models would do the
remaining job.


\section*{Acknowledgements}
%
I thank F.
Flechtner (GFZ) for helpful clarifications.
%


\end{document}